\documentclass[%
 reprint,
 superscriptaddress,
nofootinbib,
 amsmath,amssymb,
 aps,
]{revtex4-2}

\usepackage[dvipdfmx]{graphicx}
\usepackage{dcolumn}
\usepackage{bm}
\usepackage{bbm}
\usepackage[colorlinks=true, allcolors=blue]{hyperref}
\usepackage[usenames]{color}

\newcommand{\dd}{\mathrm{d}}
\newcommand{\ii}{\mathrm{i}}
\newcommand{\ee}{\mathrm{e}}
\newcommand{\GF}{G_{\mathrm{F}}}
\newcommand{\abs}[1]{\left\lvert#1\right\rvert}

\begin{document}

\title{Simple method for determining asymptotic states of fast neutrino-flavor conversion}

\author{Masamichi Zaizen}
\email{zaizen@heap.phys.waseda.ac.jp}
\affiliation{Faculty of Science and Engineering, Waseda University, Tokyo 169-8555, Japan}

\author{Hiroki Nagakura}
\affiliation{Division of Science, National Astronomical Observatory of Japan, 2-21-1 Osawa, Mitaka, Tokyo 181-8588, Japan}

\date{\today}

\begin{abstract}
Neutrino-neutrino forward scatterings potentially induce collective neutrino oscillation in dense neutrino gases in astrophysical sites such as core-collapse supernovae (CCSN) and binary neutron star mergers (BNSM).
In this paper, we present a detailed study of fast neutrino-flavor conversion (FFC), paying special attention to asymptotic states, by means of stability analysis and local simulations with a periodic boundary condition.
We find that asymptotic states can be characterized by two key properties of FFC:
(1) the conservation of lepton number for each flavor of neutrinos and (2) the disappearance of ELN(electron neutrino-lepton number)-XLN(heavy-leptonic one) angular crossings in the spatial- or time-averaged distributions.
The system which initially has the positive (negative) ELN-XLN density reaches a flavor equipartition in the negative (positive) ELN-XLN angular directions, and the other part compensates it to preserve the conservation laws.
These properties of FFCs offer an approximate scheme determining the survival probability of neutrinos in asymptotic states without solving quantum kinetic equations.
We also demonstrate that the total amount of flavor conversion can vary with species-dependent neutrino distributions for identical ELN-XLN ones.
Our results suggest that even shallow or narrow ELN angular crossings have the ability to shuffle large amount of neutrinos among flavors through FFC in the angular directions where neutrinos are more abundant, indicating the need for including the effects of FFCs in the modeling of CCSN and BNSM.
\end{abstract}

\maketitle


\section{Introduction}
A copious amount of neutrinos are emitted from inner regions of core-collapse supernovae (CCSN) and binary neutron star mergers (BNSM).
Their inner dynamics are sensitive to neutrino physics, including neutrino-matter interactions, energy/momentum/lepton transport, and flavor conversion.
Quantifying the impact of each element is one of the major tasks in developing theoretical models of CCSNe and BNSM.

Neutrinos can change their flavors during the propagation, which may affect fluid dynamics and nucleosynthesis in CCSNe and BNSM through neutrino-matter interactions.
In environments with dense neutrino media, refractive effects by neutrino self-interactions play non-negligible roles in flavor conversions or even become dominant in the neutrino oscillation Hamiltonian.
This potentially leads to collective neutrino oscillation \cite{Pantaleone:1992,Pantaleone:1992a,Sigl:1993,Duan:2010,Chakraborty:2016,Tamborra:2021,Richers:2022b}.
The oscillation frequency increases with the neutrino number density, and the wavelength can be much shorter than the scale height of the stellar structure.
In the last two decades, slow instability (or slow flavor conversion), which occurs by interacting between vacuum and self-interaction potentials, has been studied and exhibited dramatic flavor conversions \cite{Duan:2006b,Hannestad:2006,Fogli:2007,Dasgupta:2009}.
The slow flavor conversion is, however, suppressed by dense background matter due to the relatively slow oscillation scale \cite{Esteban-Pretel:2008,Chakraborty:2011,Sarikas:2012a}.
It suggests that the flavor conversion does not occur in the vicinity of neutrino spheres, and the impact on CCSN and BNSM seems subtle.
Although removing assumptions of symmetries, which have been commonly imposed in the traditional approach, alleviates the suppression, recent numerical simulations suggest that the matter suppression is still dominant in the inner region of CCSN and BNSM \cite{Raffelt:2013,Chakraborty:2014a,Mangano:2014,Dasgupta:2015,Chakraborty:2016b,Capozzi:2016,Zaizen:2021}.

Recently, fast neutrino-flavor conversion (FFC) has received significant attention.
It evolves independently from vacuum potential, and the flavor conversion is purely dictated by self-interaction potentials.
The time scale of the fast instability can be much shorter than the slow one, and therefore the large flavor conversion may happen even inside neutrino spheres \cite{Sawyer:2005,Sawyer:2009,Sawyer:2016,Chakraborty:2016a,Dasgupta:2017,Shalgar:2022a,Shalgar:2022b}.
The necessary and sufficient condition for the occurrence of FFC is equivalent to the presence of angular crossings with respect to $\left(f_{\nu_\alpha}-f_{\bar{\nu}_\alpha}\right)-\left(f_{\nu_\beta}-f_{\bar{\nu}_\beta}\right)$ \cite{Morinaga:2022,Dasgupta:2022} 
(hereafter called ELN-XLN angular crossings), where $f_{\nu_\alpha}$ and $f_{\bar{\nu}_\alpha}$ denote the distribution function of $\alpha$-type flavor neutrinos and their anti-partners, respectively.
In the context of CCSN and BNSM, the three-species treatment ($\nu_e$, $\bar{\nu}_e$, and $\nu_X$), corresponding to the condition that heavy-leptonic flavors have identical distributions, is a reasonable approximation (but see Refs.\,\cite{Bollig:2017,Fischer:2020}).
As a result, the stability condition can be essentially reduced to the presence of crossings in the electron-neutrino lepton number (ELN) angular distribution
(see, e.g., Refs.\,\cite{Izaguirre:2017,Capozzi:2017,Yi:2019,DelfanAzari:2019,Chakraborty:2020}).
This motivated ELN crossing searches in neutrino data of recent simulations.
These studies indicate that FFC likely occurs in CCSN \cite{Abbar:2019,Nagakura:2019,DelfanAzari:2020,Abbar:2020,Morinaga:2020,Glas:2020,Abbar:2020b,Capozzi:2021,Nagakura:2021b,Harada:2022,Akaho:2022}
and in BNSM \cite{Wu:2017,Li:2021,Just:2022,Richers:2022a,Grohs:2022}.
This exhibits the importance of studying the non-linear regime of FFC.

Studying flavor evolution in the non-linear phase requires directly solving a quantum kinetic equation (QKE).
At the moment, global simulations of QKE are far from achieved due to computational limitations, and therefore recent efforts for FFC have focused on local simulations \cite{Dasgupta:2018,Capozzi:2020,Johns:2020a,Shalgar:2021,Shalgar:2021b}, particularly under spatially-inhomogeneous perturbations \cite{Martin:2020a,Bhattacharyya:2020,Bhattacharyya:2021,Bhattacharyya:2022,Wu:2021,Richers:2021,Richers:2021a,Richers:2022,Zaizen:2021a,Abbar:2022,Capozzi:2022a}.
Refs.\,\cite{Bhattacharyya:2020,Bhattacharyya:2021,Bhattacharyya:2022} have presented that the system reaches a quasi-steady state with a flavor equipartition, which was concluded to be induced by flavor depolarization within the spatial domain.
Ref.\,\cite{Wu:2021} has revealed that the interactions among flavor waves develop the small-scale structures by employing several initial asymmetric number density ratios of $\bar{\nu}_e$ to $\nu_e$ and different initial perturbation.
Ref.\,\cite{Zaizen:2021a} has presented that non-linear couplings among spatial Fourier modes develop a cascade and the excited modes are transferred from linear ones according to the dispersion relation into the smaller-scale ones.
Ref.\,\cite{Richers:2021a} has performed FFC simulations with three-spatial dimensions and shown that the saturation behaviors are less affected compared to the linear growth.
All of local simulations on inhomogeneous FFC have similarly exhibited the common and attractive phenomenon, a flavor equipartition in neutrino angular distributions.
Although there has been significant progress in developing numerical simulations, our physical understanding of the non-linear dynamics of FFC is still limited.
Some key questions remain unanswered;
what ingredients characterize the quasi-steady state (see also Refs.\,\cite{Bhattacharyya:2020,Bhattacharyya:2021,Wu:2021,Bhattacharyya:2022})?;
how neutrino-matter interactions (collision term) alter dynamics of flavor conversions (see also Refs.\,\cite{Shalgar:2021a,Kato:2021,Sasaki:2022a,Kato:2022,Martin:2021,Sigl:2022,Johns:2022,Johns:2022a,Xiong:2022a})?;
is there any qualitative difference in the quasi-steady state between local- and global scales (see also Refs.\,\cite{Nagakura:2022a,Nagakura:2022b})?
Answering these questions requires a complementary analysis shining the spotlight on key quantities characterizing the non-linear dynamics of flavor conversion.

In this paper, we explore detailed features of asymptotic states of FFC by means of stability analysis and local simulations with a periodic boundary condition.
This study also provides the rationale for the results of our previous works;
flavor conversions reach a quasi-steady state in which ELN-XLN angular crossings disappear \cite{Nagakura:2022a,Nagakura:2022b}; 
how the boundary condition of the simulation box has an influence on the asymptotic state \cite{Nagakura:2022}, which is associated with a conservation law of ELN and XLN.
Based on these arguments, we provide an approximate scheme to determine the asymptotic states of FFC for a periodic boundary condition, which is in reasonable agreement with numerical simulations\footnote{In our previous paper \cite{Nagakura:2022b}, we provide the same approximate scheme but for the Dirichlet boundary condition.}.

The paper is organized as follows.
We describe two requirements determining the final fates of local FFC in terms of the stability analysis and the conservation laws in Sec.\,\ref{Sec.II:non-linear_req}.
We then present how FFC eliminates the angular crossings and establishes a quasi-steady state in Sec.\,\ref{Sec.III:inhomogeneous}.
In Sec.\,\ref{Sec.IV:comparison}, we compare our phenomenological model to previous studies, which highlights the novelty of the present study.
Finally, we conclude and discuss our study in Sec.\,\ref{Sec.V:conclusion}.

\section{Characterization of flavor evolution}\label{Sec.II:non-linear_req}
We can obtain the flavor evolution by directly calculating the non-linear QKE for the neutrino self-interactions.
At first, we will analyze what characterizes the behaviors of FFC before performing the non-linear simulation.

\subsection{Stability analysis for asymptotic state}\label{Sec.IIA:stability}
Neutrino flavor conversions can be described by a QKE for density matrices of neutrinos:
\begin{equation}
    (\partial_t+\boldsymbol{v}\cdot\nabla)\rho = -\ii\left[\mathcal{H},\rho\right] + \mathcal{C}.
\end{equation}
Here we consider the plane-parallel geometry to drop the advection term of neutrino angular directions in the momentum space.
The Hamiltonian $\mathcal{H}$ of neutrino oscillations is 
\begin{equation}
    \mathcal{H} = U\frac{M^2}{2E}U^{\dagger} + v^{\mu}\Lambda_{\mu} + \sqrt{2}\GF\int\dd\Gamma^{\prime}\, v^{\mu}v_{\mu}^{\prime} \rho^{\prime},
\end{equation}
where $\Gamma$ specifies the neutrino energy $E$ and the flight direction $\boldsymbol{v}$.
Following the same convention 
as Ref.\,\cite{Airen:2018},
 the phase-space integration is $\int\dd\Gamma = \int^{+\infty}_{-\infty}\dd E\, E^2\int\dd\boldsymbol{v}/(2\pi)^3$.
In the first term, $M^2$ and $U$ denote he mass-squared matrix and the Pontecorvo-Maki-Nakagawa-Sakata matrix, respectively.
The second term represents matter-induced oscillation, where $v^{\mu}=(1,\boldsymbol{v})$ and $\Lambda_{\mu} = \sqrt{2}\GF\,\mathrm{diag}[\{j_{\mu}^{\ell}\}]$.
In the study, we ignore the collision term for simplicity.

For an effective two-flavor scenario, we can decompose the density matrix at a certain space-time position $(t,\boldsymbol{x})$ into the trace- and traceless parts as,
\begin{equation}
    \rho(t,\boldsymbol{x}) = \frac{\mathrm{Tr}\rho}{2}\mathbbm{1}_2 + \frac{f_{\nu_e}-f_{\nu_x}}{2}
    \begin{pmatrix}
    s & S \\ S^* & -s
    \end{pmatrix},
    \label{eq:deco}
\end{equation}
where $s$ is a real number, $S$ is a complex one, and $f_{\nu_{\alpha}}$ is an occupation number density for a flavor $\alpha$. 
We note that time dependence of the traceless part (the second term in the right-hand side of Eq.\,\eqref{eq:deco}) is handled by $s$ and $S$ in our method, indicating that the prefactor ($(f_{\nu_e}-f_{\nu_x})/2$) can be determined arbitrarily.
In this study, we adopt spatial-averaged distributions at the beginning of each simulation ($t=0$).
On the other hand, all time- and spatial dependence in the trace part (the first term in the right-hand side of Eq.\,\eqref{eq:deco}) is handled by $\mathrm{Tr\rho}$ (which does not affect flavor conversion, though).
We define a spatial-averaged spectral difference between $\nu_e$ and $\nu_x$ at $t=0$ as
\begin{align}
    g_{E,\boldsymbol{v}} = 
    \begin{cases}
        f_{\nu_e} - f_{\nu_x} \,\,\,\mathrm{for\,\,} E>0 \\
        f_{\bar{\nu}_x} - f_{\bar{\nu}_e} \,\,\,\mathrm{for\,\,} E<0
    \end{cases}.
\end{align}
By using these variables, the governing equation for $S$ (the off-diagonal component of the traceless part) can be written as,
\begin{align}
\ii v^{\mu}\partial_{\mu}S_{E,\boldsymbol{v}} 
    &\,= -\omega_{\mathrm{V}}\cos 2\theta_{\mathrm{V}}\, S_{E,\boldsymbol{v}} - \omega_{\mathrm{V}}\sin 2\theta_{\mathrm{V}}\, s_{E,\boldsymbol{v}} \notag\\
    &\,\,+ v^{\mu}\left(\Lambda^{11}_{\mu}-\Lambda^{22}_{\mu}\right)S_{E,\boldsymbol{v}} \notag\\
    &\,\,+ S_{E,\boldsymbol{v}}\, v^{\mu}\int\dd\Gamma^{\prime}\, v_{\mu}^{\prime}\, g_{E^{\prime},\boldsymbol{v}^{\prime}} s_{E^{\prime},\boldsymbol{v}^{\prime}} \notag\\
    &\,\,- s_{E,\boldsymbol{v}}\, v^{\mu}\int\dd\Gamma^{\prime}\, v_{\mu}^{\prime}\, g_{E^{\prime},\boldsymbol{v}^{\prime}} S_{E^{\prime},\boldsymbol{v}^{\prime}},
    \label{eq:CapiSv_eq}
\end{align}
while
\begin{align}
\ii v^{\mu}\partial_{\mu}s_{E,\boldsymbol{v}}
    &\,= \frac{\omega_{\mathrm{V}}}{2}\sin 2\theta_{\mathrm{V}} \left(S_{E,\boldsymbol{v}}^{*}-S_{E,\boldsymbol{v}}\right) \notag\\
    &\,\,+\frac{1}{2} S_{E,\boldsymbol{v}}^{*}\, v^{\mu}\int\dd\Gamma^{\prime}\, v_{\mu}^{\prime}\, g_{E^{\prime},\boldsymbol{v}^{\prime}} S_{E^{\prime},\boldsymbol{v}^{\prime}} \notag\\
    &\,\,-\frac{1}{2} S_{E,\boldsymbol{v}}\, v^{\mu}\int\dd\Gamma^{\prime}\, v_{\mu}^{\prime}\,g_{E^{\prime},\boldsymbol{v}^{\prime}} S_{E^{\prime},\boldsymbol{v}^{\prime}}^{*}
    \label{eq:sv_eq}
\end{align}
is the equation for $s$ (the diagonal component).
In the expression, $\omega_{\mathrm{V}}$ and $\theta_{\mathrm{V}}$ denote a vacuum frequency and a mixing angle, respectively.
It should be stressed that any approximations are not imposed in deriving Eqs.\,\eqref{eq:CapiSv_eq} and \eqref{eq:sv_eq} from QKE.

Below, let us prepare the stability analysis in the non-linear phase.  
In the linear regime (under the choice of fixed points as flavor eigenstates), the diagonal term is given by $s = \sqrt{1-\abs{S}^2}\simeq 1$, which can be directly inserted to Eq.\,\eqref{eq:CapiSv_eq}, and then we can obtain the dispersion relation for $S$ with the plane wave ansatz.
In the non-linear regime, however, $s$ would substantially deviate from unity, and more importantly, it depends on time and space, which causes mode couplings (in other words, we need to compute the convolutions of $S$ with $s$, see Eq.\,\eqref{eq:QKE_fft}).
It should also be mentioned that the stability needs to be determined globally, if the background component is not uniform.
These are main obstacles to deriving dispersion relation for $S$ in the non-linear regime.

In our prescription, we drop the mode coupling by assuming that $s$ is constant (but not unity).
One may choose $s$ as that at $(t,\boldsymbol{x})$, i.e., the space-time location where the stability analysis is conducted.
Another prescription is to adopt the spacial- or time-averaged quantity.
In fact, we are interested in the overall trend of non-linear saturation and quasi-steady state of flavor conversion, which can be characterized by those averaged distributions.
In the following, we leave the notation as $s$, but need to keep in mind this assumption.

We adopt the plane wave ansatz to $S$,
\begin{equation}
    S_{E,\boldsymbol{v}}(t,\boldsymbol{x}) \propto Q_{E,\boldsymbol{v}}(\Omega,\boldsymbol{K})\,\ee^{-\ii(\Omega t - \boldsymbol{K}\cdot\boldsymbol{x})},
\end{equation}
and then the governing equation for $S_{E,\boldsymbol{v}}$ can be recast into
\begin{align}
v^{\mu}k_{\mu}Q_{E,\boldsymbol{v}}
    = -\omega_{\mathrm{V}} Q_{E,\boldsymbol{v}} - s_{E,\boldsymbol{v}}\, v^{\mu}\int\dd\Gamma^{\prime}\, v_{\mu}^{\prime}\, g_{E^{\prime},\boldsymbol{v}^{\prime}} Q_{E^{\prime},\boldsymbol{v}^{\prime}},
\end{align}
where $k_{\mu}\equiv (\omega,\boldsymbol{k}) = K_{\mu}-\Lambda^{ex}_{\mu}-\Phi^{ex}_{\mu}$ with $\Lambda^{ex} \equiv \Lambda^{11}-\Lambda^{22}$ and $\Phi^{ex}_{\mu} = \int\dd\Gamma\, v_{\mu}\, g_{E,\boldsymbol{v}} s_{E,\boldsymbol{v}}$.
In a co-rotating frame where the vacuum term oscillates quickly and the off-diagonal terms are averaged to zero by the matter term, the mixing angle $\theta_{\mathrm{V}}$ can be set to zero.
As a result, the nontrivial solutions for $Q_{E,\boldsymbol{v}}$ are given by, using the metric $\eta=\mathrm{diag}(+,-,-,-)$,
\begin{equation}
    D(\omega,\boldsymbol{k}) \equiv \mathrm{det}\left[\Pi^{\mu\nu}(k)\right] = 0,
\end{equation}
where
\begin{equation}
    \Pi^{\mu\nu} = \eta^{\mu\nu} + \int\dd\Gamma\,\,g_{E,\boldsymbol{v}}s_{E,\boldsymbol{v}}\frac{v^{\mu}v^{\nu}}{v^{\lambda}k_{\lambda}+\omega_{\mathrm{V}}}.
    \label{eq:Piorig}
\end{equation}
We note that $\omega_{\mathrm{V}}$ dependence accounts for the slow flavor conversion.
It may be useful to rewrite Eq.\,\eqref{eq:Piorig} as
\begin{equation}
    \Pi^{\mu\nu} = \eta^{\mu\nu} + \int\dd\Gamma\,\,g^{ex}_{E,\boldsymbol{v}}(t,\boldsymbol{x})\frac{v^{\mu}v^{\nu}}{v^{\lambda}k_{\lambda}+\omega_{\mathrm{V}}},
    \label{eq:Pi2nd}
\end{equation}
where
\begin{align}
g^{ex}_{E,\boldsymbol{v}}(t,\boldsymbol{x}) &= (P_{ee}\,g^{e}_{E,\boldsymbol{v}}+P_{xe}\,g^{x}_{E,\boldsymbol{v}}) - (P_{ex}\,g^{e}_{E,\boldsymbol{v}}+P_{xx}\,g^{x}_{E,\boldsymbol{v}}) \notag\\
    &= (P_{ee}-P_{ex})\,g_{E,\boldsymbol{v}} \notag\\
    &= g_{E,\boldsymbol{v}}\,s_{E,\boldsymbol{v}}(t,\boldsymbol{x}).
\end{align}
In the expression, we use the relation of $P_{ex} = (1-s_{E,\boldsymbol{v}})/2$, where $P_{\alpha \beta}$ is defined as a transition probability from $\alpha$-type flavor at the initial state to $\beta$-type one at $t$.
If there is a nonzero imaginary part in $k_{\mu}$, the flavor instability $S_{E,\boldsymbol{v}}$ can grow or dump in space and time.

FFC can be driven only by the neutrino self-interactions, and we can ignore the vacuum term.
In the fast limit, $\omega_{\mathrm{V}}=0$, the density matrix loses the explicit energy-dependence in the QKE, and therefore Eq.\,\eqref{eq:Pi2nd} can be more concisely rewritten.
We define an $\alpha$-flavor lepton number ($\alpha$LN) angular distribution
\begin{equation}
    G_{\boldsymbol{v}}^{\alpha} = \sqrt{2}\GF\int^{\infty}_{0}\frac{\dd E\,E^2}{2\pi^2}\left[f_{\nu_{\alpha}}(E,\boldsymbol{v})-f_{\bar{\nu}_{\alpha}}(E,\boldsymbol{v})\right]
\end{equation}
and the difference $G^{\alpha\beta}_{\boldsymbol{v}} \equiv G^{\alpha}_{\boldsymbol{v}}-G^{\beta}_{\boldsymbol{v}}$ between $\alpha$ and $\beta$ flavors.
The dispersion relation only for fast instability is reduced to
\begin{equation}
    D(\omega,\boldsymbol{k}) \equiv \mathrm{det}\left[\Pi^{\mu\nu}(k)\right] = 0,
    \label{eq:DR}
\end{equation}
where
\begin{equation}
    \Pi^{\mu\nu} = \eta^{\mu\nu} + \int\frac{\dd\boldsymbol{v}}{4\pi}\,\,G^{ex}_{\boldsymbol{v}}(t,\boldsymbol{x})\frac{v^{\mu}v^{\nu}}{v^{\lambda}k_{\lambda}}.
    \label{eq:DR_Gv}
\end{equation}
For the stability with respect to the spatial- or time-averaged distributions, we use the averaged quantity of $G^{ex}_{\boldsymbol{v}}(t,\boldsymbol{x})$ in Eq.\,\eqref{eq:DR_Gv}.

A few remarks should be made here.
The dispersion relation is essentially the same as that used in linear stability analysis.
In fact, the growth (or damp) of $S$ is determined solely by the diagonal components of the density matrix.
Based on the same argument made by Ref.\,\cite{Morinaga:2022}, the stability can be determined by the presence of ELN-XLN crossings even in the non-linear phase\footnote{
However, a homogeneous mode can be stable even if there exists ELN-XLN crossing.
This is simply because unstable solutions possibly appear only in inhomogeneous modes.
Nevertheless, ELN-XLN angular distributions are key quantities to characterize the stability of FFC.},
suggesting that ELN-XLN angular distributions are fundamental quantities to characterize the dynamics of flavor conversion.
On the other hand, our method ignores the mode couplings with the inhomogeneity of ELN-XLN angular distributions.
We will validate this condition by numerical simulations (see Sec.\,\ref{Sec.III:inhomogeneous}).

\subsection{Conservation laws}\label{Sec.IIB:conserv}
In the fast limit, the QKE can be decomposed into commutators with the neutrino density $H_E$ and the flux $H_F$ as
\begin{align}
    \ii(\partial_t + v_z\partial_z)\rho_{v} &= \left[H_{\nu\nu},\, \rho_{v}\right] \notag\\
    &= \left[H_E,\, \rho_{v}\right] - \left[v_z H_F,\, \rho_{v}\right],
\end{align}
where
\begin{align}
    H_E &= \sqrt{2}\GF\int\dd\Gamma^{\prime}\, \rho_{v^{\prime}} \\
    H_F &= \sqrt{2}\GF\, \int\dd\Gamma^{\prime}\, v_z^{\prime}\rho_{v^{\prime}}.
\end{align}
Moreover, integrating the QKE over the phase space, we can obtain
\begin{align}
    \partial_t H_E+ \partial_z H_F = 0,
    \label{eq:QKEmom}
\end{align}
which has a conservative form without source terms.
Note that the neutrino density is not locally conserved due to the advection (flux) term.
Both ELN and XLN are conserved in terms of spatially integrated quantity because the neutrino flux on the boundary surface is closed in the periodic case.
Therefore, flavor conversion proceeds with satisfying the ELN and XLN conservation and eventually reaches a non-linear saturation.
In other words, FFC transfers the number density from the positive (negative) part to the negative (positive) and saturates when the angular distributions are wholly positive (negative) or zero.
These properties offer a simple analytic prescription to determine an asymptotic state of FFC, which will be discussed in the next subsection.

Here, let us make an important remark.
The spatial-integrated $H_{E}$ is, in general, not conserved if another boundary condition is imposed.
This implies that the system evolves towards a different asymptotic state;
in fact we observed that the system settled into a qualitatively different quasi-steady state for the Dirichlet boundary condition, see Refs.\,\cite{Nagakura:2022,Nagakura:2022b}.

\subsection{Asymptotic states}\label{Sec.IIC:asymptotic}
From the above discussion, we can predict the asymptotic states from two requirements: the disappearance of ELN-XLN angular crossings and the conservation of ELN and XLN.
The disappearance of ELN-XLN angular crossings is equivalent to the achievement of flavor equipartition in some angular regions, which is determined as follows\footnote{In this study, we assume that there is a single ELN-XLN crossing in initial angular profiles. The study of multiple ELN crossings is currently underway and will be reported in a forthcoming paper.}.
If the total number density, $\int\dd v\, (\mathrm{ELN}-\mathrm{XLN}$), is positive, flavor conversion proceeds until the achievement of flavor equipartition on the angular parts with negative lepton number.
In the other angular part, the ELN and XLN are adjusted so as to satisfy the number conservation of ELN and XLN.
If the total number density is negative, flavor equipartition is achieved on the positive part, and then the conservation of ELN and XLN is adjusted in the negative one.
Note that if the total number density of ELN-XLN is zero, complete flavor equipartition is established in the entire angular distribution, which is consistent with the result in \cite{Wu:2021}.

Following the above consideration, we develop an approximate scheme.
The negative ELN-XLN part $A$ and positive one $B$ are defined as
\begin{align}
    A &\equiv \abs{\int_{G^{ex}_{\boldsymbol{v}}<0}\frac{\dd\boldsymbol{v}}{4\pi}G^{ex}_{\boldsymbol{v}}} \\
    B &\equiv \abs{\int_{G^{ex}_{\boldsymbol{v}}>0}\frac{\dd\boldsymbol{v}}{4\pi}G^{ex}_{\boldsymbol{v}}}.
\end{align}
When the total number density of ELN-XLN is positive, $B-A>0$, the number density corresponding to $A/2$ is transferred from the ELN to the XLN in the negative ELN-XLN directions to establish a flavor equipartition.
On the other hand, the same amount of number density is converted from the positive ELN-XLN parts to sustain the conservation for ELN (and XLN).
Consequently, in the positive part, the number density $A/2$ is subtracted from the positive number density $B$ and distributed into the XLN.
Thereby, the survival probability of electron-type neutrinos can be analytically estimated as the following simple box-like formulas:
\begin{align}
    P_{ee} &= 
    \begin{cases}
        p \,\,\,&\mathrm{for}\,\,G_v<0 \\
        1-\left(1-p\right)\dfrac{A}{B} \,\,\,&\mathrm{for}\,\,G_v>0
    \end{cases},
    \label{eq:surv_approx1}
\end{align}
where $p$ is a survival probability in the negative ELN-XLN part and becomes $p=p_{\mathrm{eq}}$, where $p_{\mathrm{eq}}$ represents the survival probability for flavor equipartition, to eliminate the ELN-XLN crossing.
Note that $p_{\mathrm{eq}}$ is $1/2$ for two-flavor of neutrinos (see below for the case with a three-flavor framework).
For the negative case, $B-A<0$, considering the opposite, we obtain
\begin{align}
    P_{ee} &= 
    \begin{cases}
        p \,\,\,&\mathrm{for}\,\,G_v>0 \\
        1-\left(1-p\right)\dfrac{B}{A} \,\,\,&\mathrm{for}\,\,G_v<0
    \end{cases}.
    \label{eq:surv_approx2}
\end{align}
In the symmetric flavor case $A=B$, our analytical scheme provides full flavor equipartition in the entire angular directions.
Note that within a three-flavor framework, we need to consider flavor equipartitions on ELN-MuLN, ELN-TauLN, and MuLN-TauLN simultaneously if they possess crossings.
In the case that MuLN is identical to TauLN, which is a reasonable condition in CCSNe and BNSM, ELN-MuLN and ELN-TauLN should reach the identical flavor equilibrium \cite{Chakraborty:2020}.
This exhibits that the non-zero ELN at the initial state is equally distributed to MuLN and TauLN, indicating that the survival probability for the case of flavor equipartion should be $p_{\mathrm{eq}}=1/3$ in three-flavor framework.

There is a caveat in the approximate scheme, however.
The box-like structure in angular distributions of $p$ would not be realistic.
In fact, we will show in Sec.\,\ref{Sec.III:inhomogeneous} that $p$ does not change discontinuously at the crossing point but rather has a smooth profile.
Nevertheless, the overall trend in asymptotic states of FFCs can be captured by the simple structure, and we leave its improvement to future work.

\section{Numerical tests}\label{Sec.III:inhomogeneous}
We perform numerical simulations on a periodic boundary condition, which validates our approach to determine asymptotic states of FFC.
We assume a one-dimensional simulation box with axial-symmetry around the z-axis.
As an initial state, we use the following ELN angular distribution, employed as $G_{4b}$ in Ref.\,\cite{Martin:2020a},
\begin{equation}
G_v^{e} = \mu\left[g_{\nu_e}(v) - \alpha g_{\bar{\nu}_e}(v)\right],
\end{equation}
where $\mu=\sqrt{2}\GF n_{\nu_e}$, $\alpha=0.92$ is an asymmetry parameter, and normalized angular distribution $g_{\nu}$ is defined as
\begin{equation}
g_{\nu} \propto \exp\left[-(v-1)^2/2\xi_{\nu}^2\right] \label{eq:g_nu}
\end{equation}
with $\xi_{\nu_e}=0.6$ and $\xi_{\bar{\nu}_e}=0.53$.
We also assume that XLN is initially zero.
Then, the ELN(-XLN) angular crossing is located at $v_c = 0.68$ and the ELN(-XLN) number density, $1-\alpha$, is positive.
Hence, we can predict that FFC proceeds so as to fill in the negative lepton number on more forward-directional side $v\gtrsim v_c$ relative to the angular crossing from the above requirements.

\begin{figure}[t]
    \centering
    \includegraphics[width=1.0\linewidth]{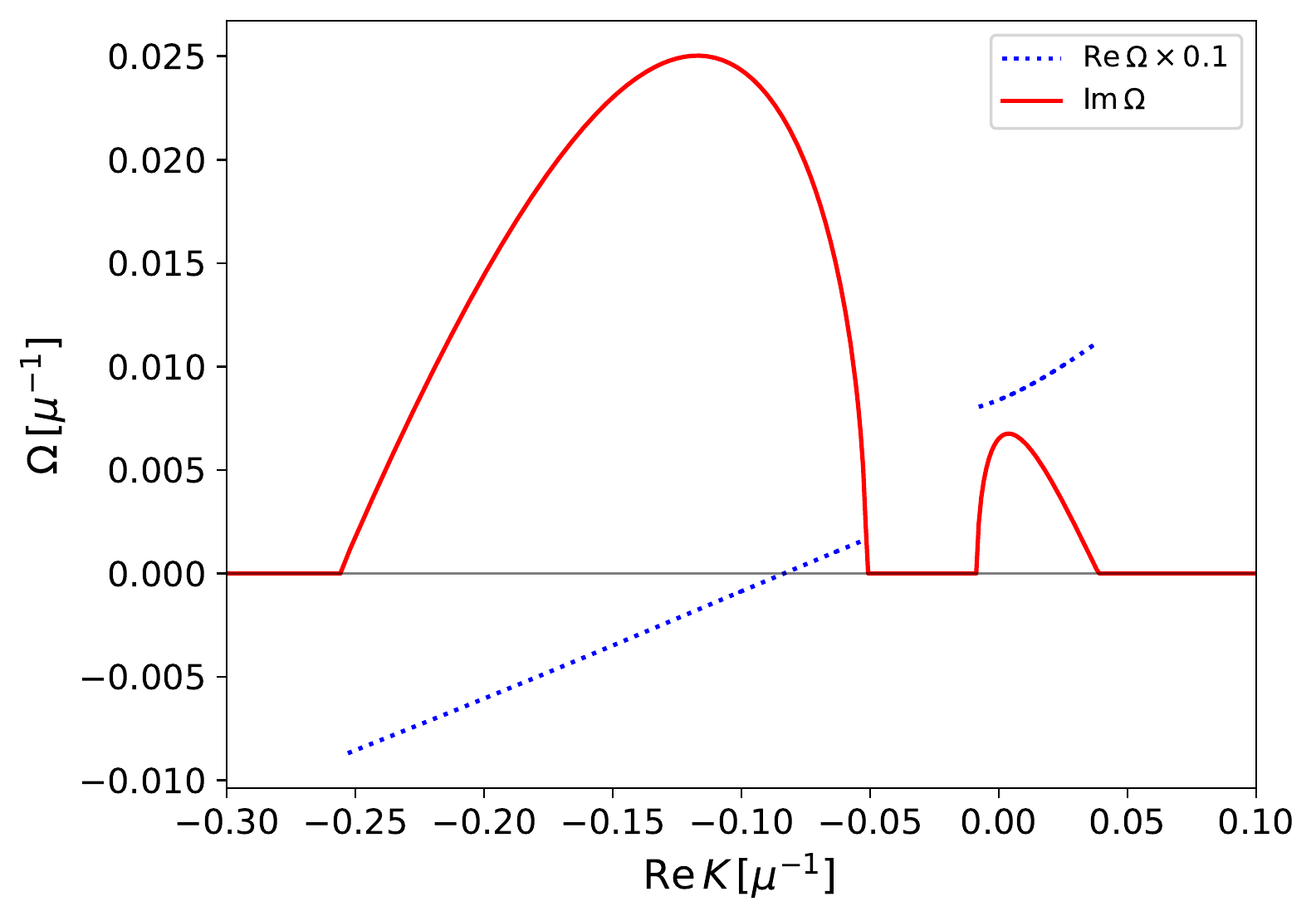}
    \includegraphics[width=1.0\linewidth]{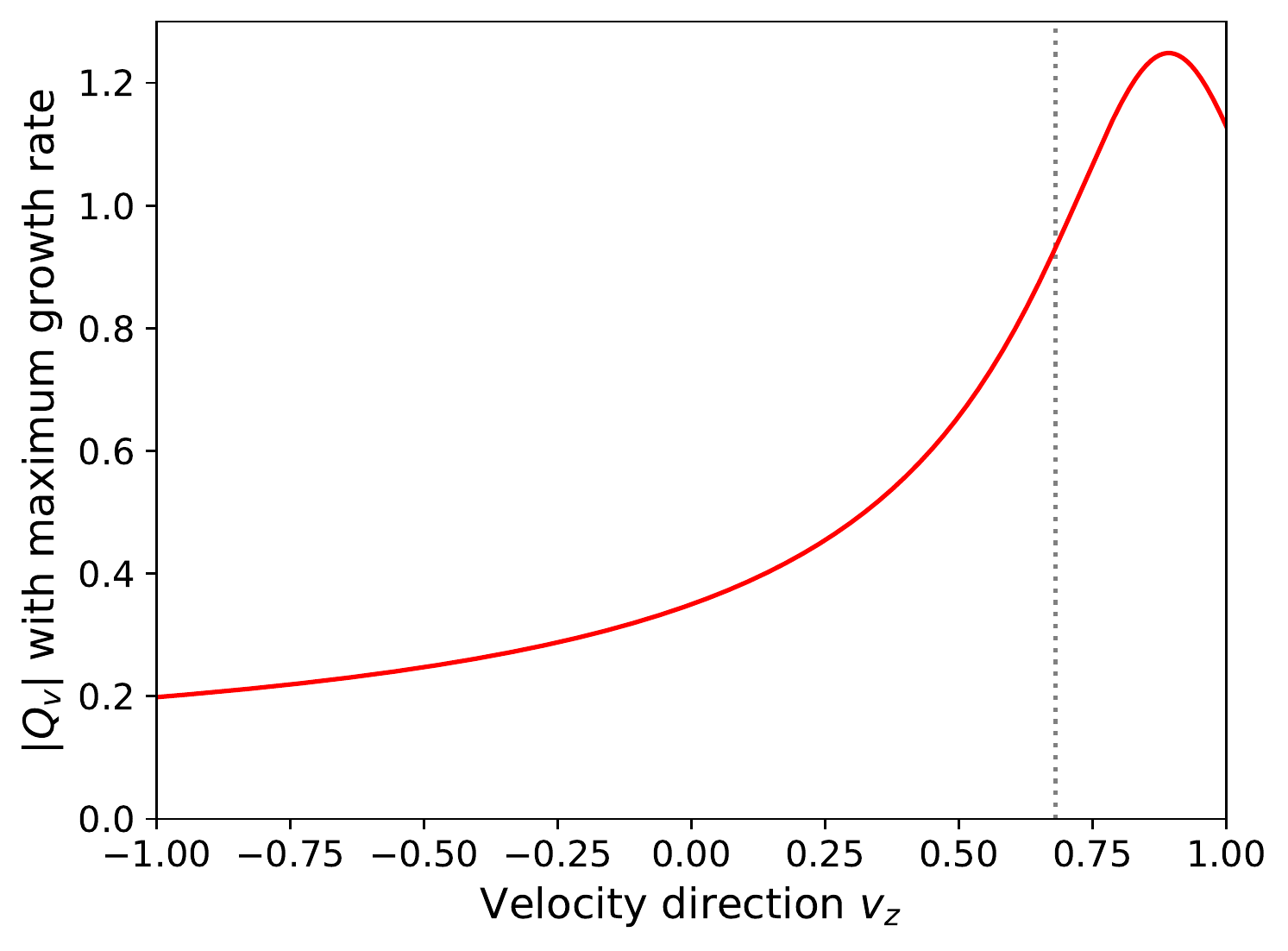}
    \caption{
    Top: Dispersion relation of $\Omega(K)$ including unstable fast modes for ELN angular distribution $G_{4b}$.
    Solid (dotted) lines are for imaginary (real) parts of $\Omega$.
    Bottom: Normalized amplitude $\abs{Q_v}$ with a maximum growth rate in the top panel.
    Vertical dotted line is a crossing point, and we find that the peak is located within the negative ELN directions.
    }
    \label{fig:DR}
\end{figure}

Figure\,\ref{fig:DR} shows the dispersion relation $\Omega(K)$ satisfying Eq.\,\eqref{eq:DR} for the initial ELN angular distribution $G_{4b}$ in the linear regime.
Spatial modes $K$ with nonzero imaginary parts $\mathrm{Im}\,\Omega$ correspond to unstable branches and can induce FFC in the non-linear regime.
In the bottom panel, we present the normalized eigenvector $\abs{Q_v}$ with the maximum growth rate in the top panel and the peak amplitude is within the negative ELN directions, $v_z > v_{c}$.
Consequently, the initial perturbation on the flavor coherent $S$ evolves strongly within the crossing in the linear regime, and then flavor conversion can occur mainly in the angular directions in a non-linear phase.
This picture agrees with the above prediction for the non-linear saturation.
The behaviors of FFC in the non-linear regime will be demonstrated in Sec.\,\ref{Sec.IIIB:EXLN}.

\subsection{Setups}
In the fast limit, neutrino density matrices lose the explicit energy-dependence in the QKE.
Hence, antineutrinos are identified to those for neutrinos, i.e., $\bar{\rho}_v=\rho_v$.
And we adopt a pseudo-spectral method using Fast Fourier Transformation implemented in the \texttt{FFTW3} library\footnote{Fastest Fourier Transform in the West, http://www.fftw.org.} to handle the spatial advection operator in the QKE (see Refs.\,\cite{Zaizen:2021a,Richers:2022} on the detail and recent applications).
Then, QKE for the spatial Fourier components in the polarization vector configuration $\rho \propto \boldsymbol{P}\cdot\boldsymbol{\sigma}$ is recast into
\begin{align}
    \partial_t\tilde{\boldsymbol{P}}^{K}_{v} &= -\mathrm{i}vK\tilde{\boldsymbol{P}}^{K}_{v} \notag\\
    &\,\,+ \sum_{K^{\prime}}\left[\int\dd v^{\prime}(1-vv^{\prime})G^e_{v^{\prime}} \tilde{\boldsymbol{P}}^{K-K^{\prime}}_{v^{\prime}}\times \tilde{\boldsymbol{P}}^{K^{\prime}}_{v}\right].
    \label{eq:QKE_fft}
\end{align}
This equation includes the non-linear coupling term among all spatial modes $K$ derived from the neutrino self-interactions.
We initially put artificial spatially-inhomogeneous perturbations on the off-diagonal components of the density matrix to trigger flavor conversion instead of nonzero vacuum mixing.
The spatial perturbation is
\begin{align}
    P_1(t=0,z,v_z) &= \epsilon(z) \\
    P_3(t=0,z,v_z) &= \sqrt{1-\epsilon(z)^2},
\end{align}
where $\epsilon(z)$ is randomly arranged between $0$ and $10^{-6}$.
The third component $P_3$ corresponds to the diagonal part of the density matrix and the decrease from unity implies the occurrence of a flavor transformation.
In performing the flavor evolution in Eq.\,\eqref{eq:QKE_fft}, once the initial conditions are built on the configuration space, they are converted into the Fourier space.

The self-interaction strength $\mu$ is a unique dimensional quantity in the QKE on the fast limit.
Hence, we can measure the time and space with the unit of $\mu^{-1}$.
Then, we adopt a one-dimensional simulation box of $L_z = 1000$ spanned by a uniform grid $N_z = 10000$ with the periodic boundary condition.
And we set $256$ angular bins on the root of Legendre polynomials and use the Gauss-Legendre quadrature for the angular integration.
We evolve Eq.\,\eqref{eq:QKE_fft} using the fourth-ordered Runge-Kutta scheme with a fixed time step size of $\Delta t = C_{\mathrm{CFL}}\Delta z$, where the Courant-Friedrichs-Lewy number $C_{\mathrm{CFL}}=0.4$ as in Ref.\,\cite{Wu:2021}.

\subsection{ELN-XLN analysis}\label{Sec.IIIB:EXLN}
We perform the inhomogeneous modeling of local FFC with periodic boundary.
In the fast limit, flavor conversion is determined only by the ELN-XLN angular distributions.
Here, we focus on the time evolution of FFC in terms of ELN and XLN.

\begin{figure}[t]
    \centering
    \includegraphics[width=0.95\linewidth]{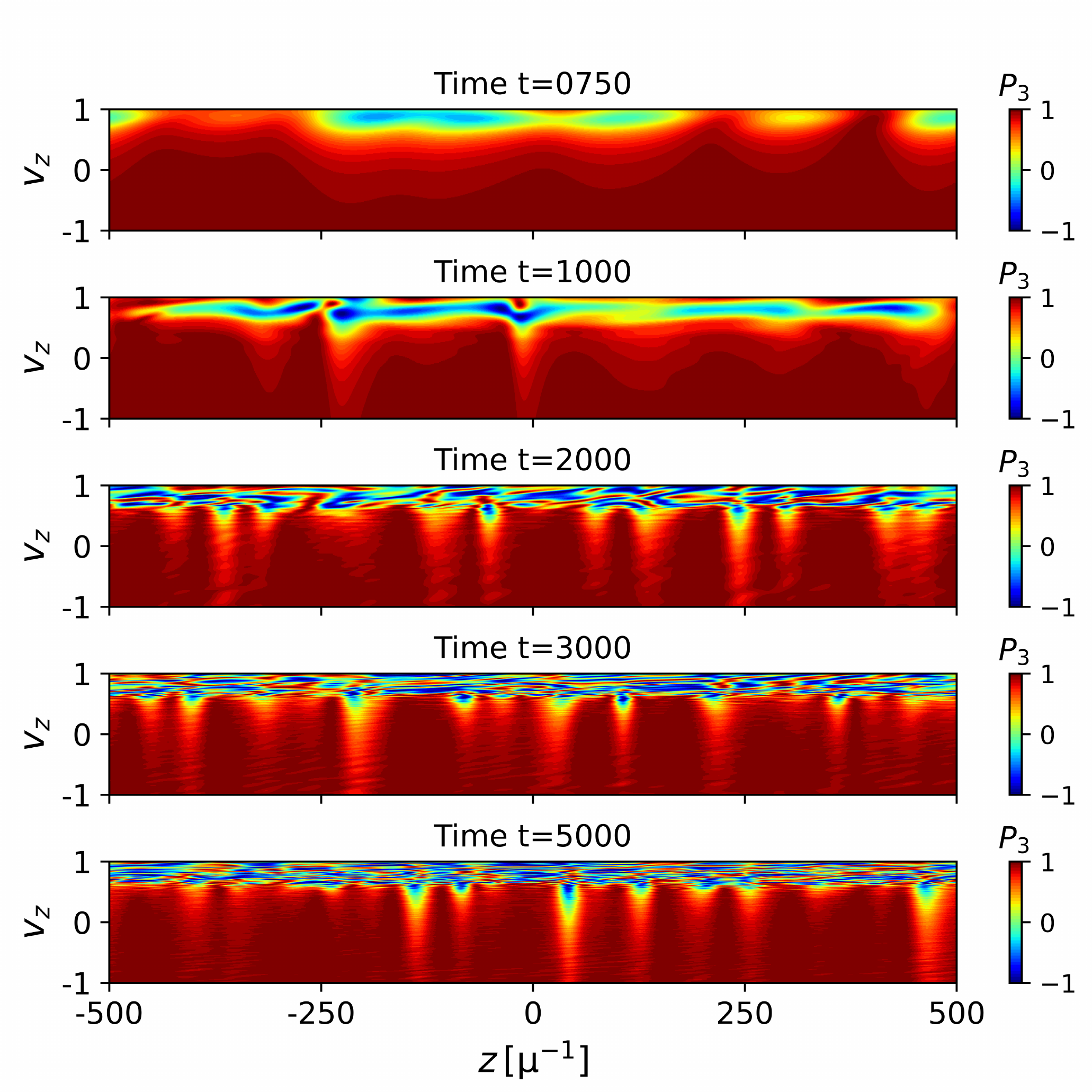}
    \caption{
    $P_3(t,z,v_z)$ on the space-velocity $(z-v_z)$ plane at five time snapshots $t=750,1000,2000,3000$, and $5000$.
    }
    \label{fig:inhomogeneous_P3}
\end{figure}

Figure\,\ref{fig:inhomogeneous_P3} shows $P_3(t,z,v_z)$ on the space-velocity $(z-v_z)$ plane at five representative time snapshots $t=750,1000,2000,3000$, and $5000$.
Flavor conversions emerge everywhere in space but mainly in $v_{z} \gtrsim 0.7$ for the angular direction.
Once the flavor conversion enters into a non-linear phase, flavor waves interfere with each other mainly due to the spatial advection and then cultivate smaller-scale structures.
After $t=2000$, the entire system establishes a quasi-steady state with temporal and spatial variations.

\begin{figure}[t]
    \centering
    \includegraphics[width=1.0\linewidth]{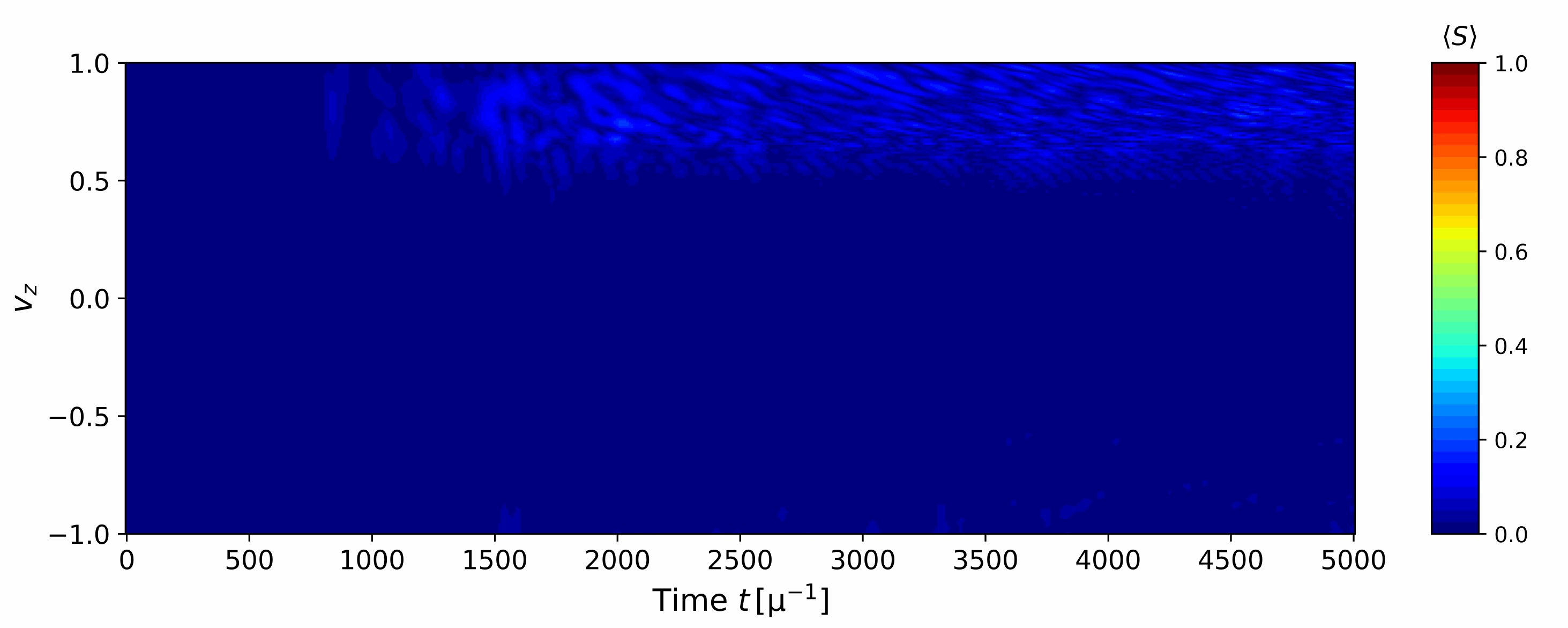}
    \includegraphics[width=1.0\linewidth]{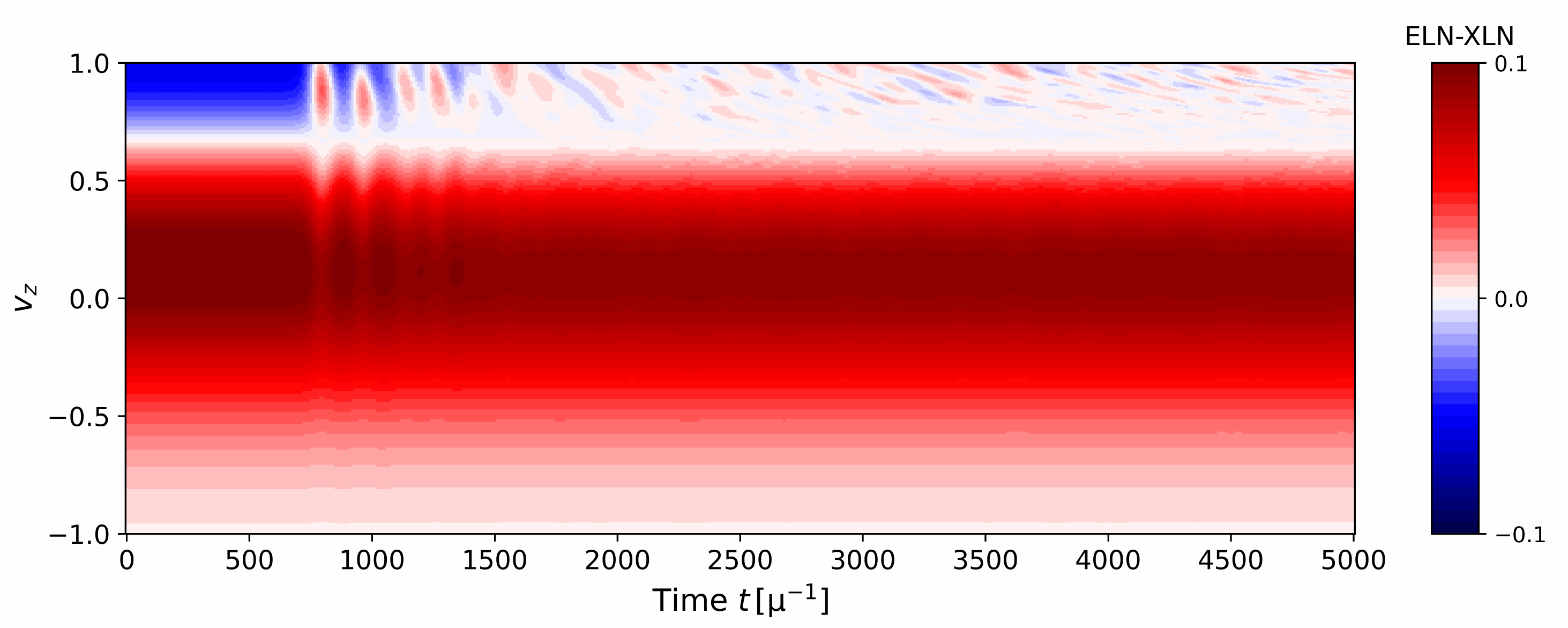}
    \includegraphics[width=0.95\linewidth]{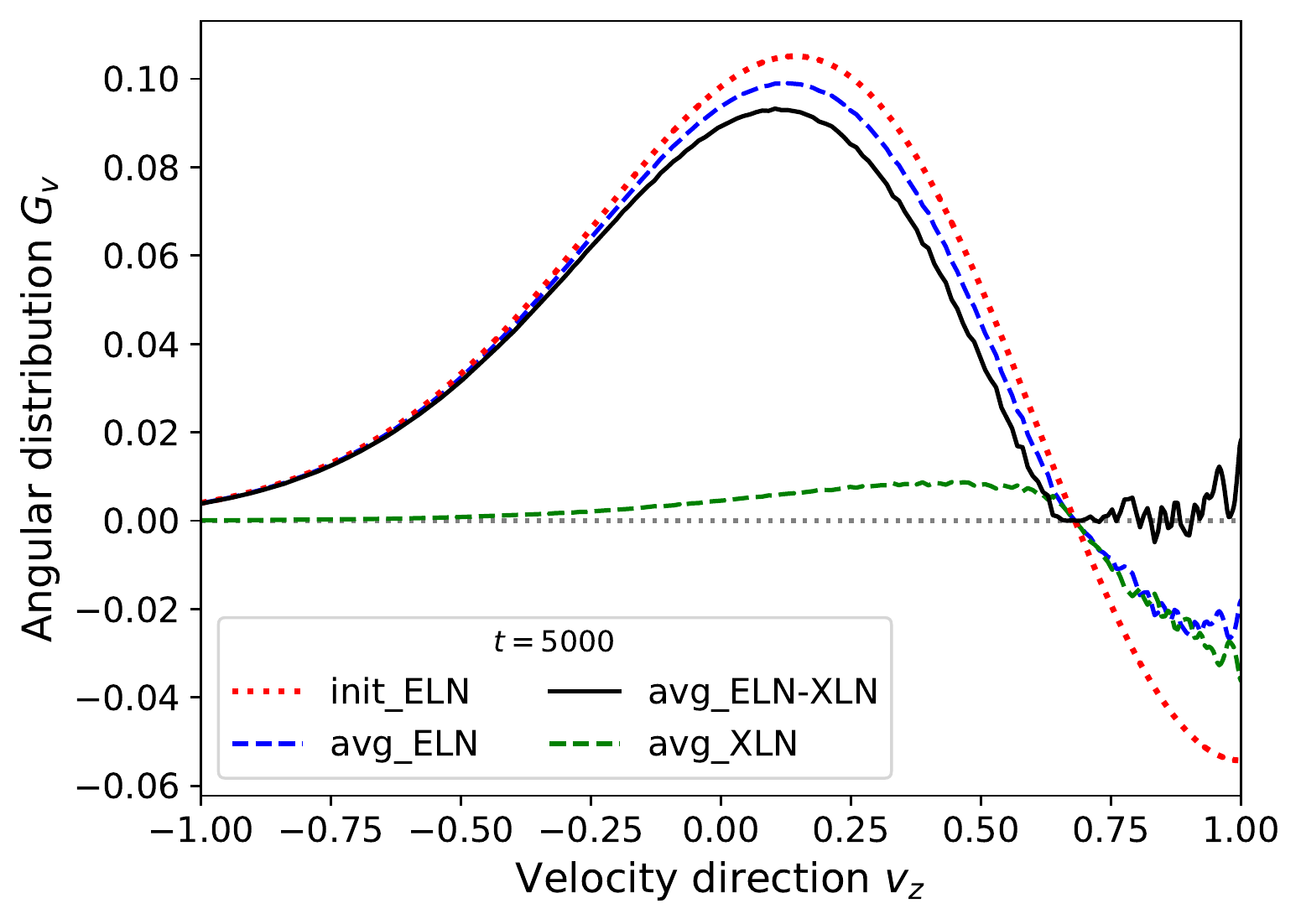}
    \caption{
    Top: Spatial-averaged flavor coherent $\langle S\rangle$ on the $t-v_z$ plane.
    Middle: Spatial-averaged ELN-XLN angular distributions on the $t-v_z$ plane.
    Bottom: Spatial-averaged angular distributions for some important quantities at $t=5000$.
    In the bottom panel, red dotted line is an initial ELN angular distribution.
    Blue (green) dashed line is an ELN (XLN) one.
    Black solid line corresponds to an ELN-XLN angular distribution, which exhibits that the angular crossing almost disappears.
    }
    \label{fig:inhomogeneous_sp_avg}
\end{figure}

To capture the overall trend of the quasi-steady state, we adopt the spatial-averaged $G^{ex}_{\boldsymbol{v}}(t,\boldsymbol{x})$ in computing the dispersion relation.
As we have already mentioned, the stability can be determined by the ELN-XLN angular crossing; hence, we focus on ELN-XLN distributions below.
Figure\,\ref{fig:inhomogeneous_sp_avg} portrays spatial-averaged flavor coherent $\langle S\rangle$ and ELN-XLN distributions as a function of time and neutrino angle in the top and middle panels, meanwhile we also display some relevant quantities of angular distribution at $t=5000$ in the bottom one.
As shown in the top panel, the spatial-averaged flavor coherent $\langle S\rangle$ is clearly small even for the late phases and is $\lesssim 0.1$, and then the neutrino density matrix is almost in the flavor state.
As shown in the middle panel, the ELN-XLN in $v_z\gtrsim 0.7$ approaches to zero with time (transiting from blue to white), meanwhile it has a subtle change in other angular directions.
In the early phase when flavor conversion starts to occur, large-scale coherent oscillation appears, but it cascades to small scale with decreasing the amplitude.
This exhibits that the system evolves towards a quasi-steady state with damping the ELN-XLN angular crossings.
Assuming that $s$ is constant in the quasi-steady state based on the two panels, the stability of FFC can be evaluated through the spatially-averaged ELN-XLN angular distributions.
We can delve into the detailed angular distributions in the quasi-steady state, at $t=5000$, in the bottom panel of Fig.\,\ref{fig:inhomogeneous_sp_avg}.
As shown in the panel, the ELN angular distribution still has a crossing even after the non-linear saturation.
However, XLN is no longer zero due to flavor conversion, and we find that it compensates the angular distribution so as to make ELN-XLN be zero in $v_z\gtrsim 0.7$.
This is the reason why further flavor conversion does not occur in the spatially averaged domain in our simulation, which is consistent with our claim obtained by the stability analysis.
As a result, the spatial-averaged ELN-XLN angular distribution is entirely positive or zero, and does not have any crossings in all directions.

\begin{figure}[t]
    \centering
    \includegraphics[width=1.0\linewidth]{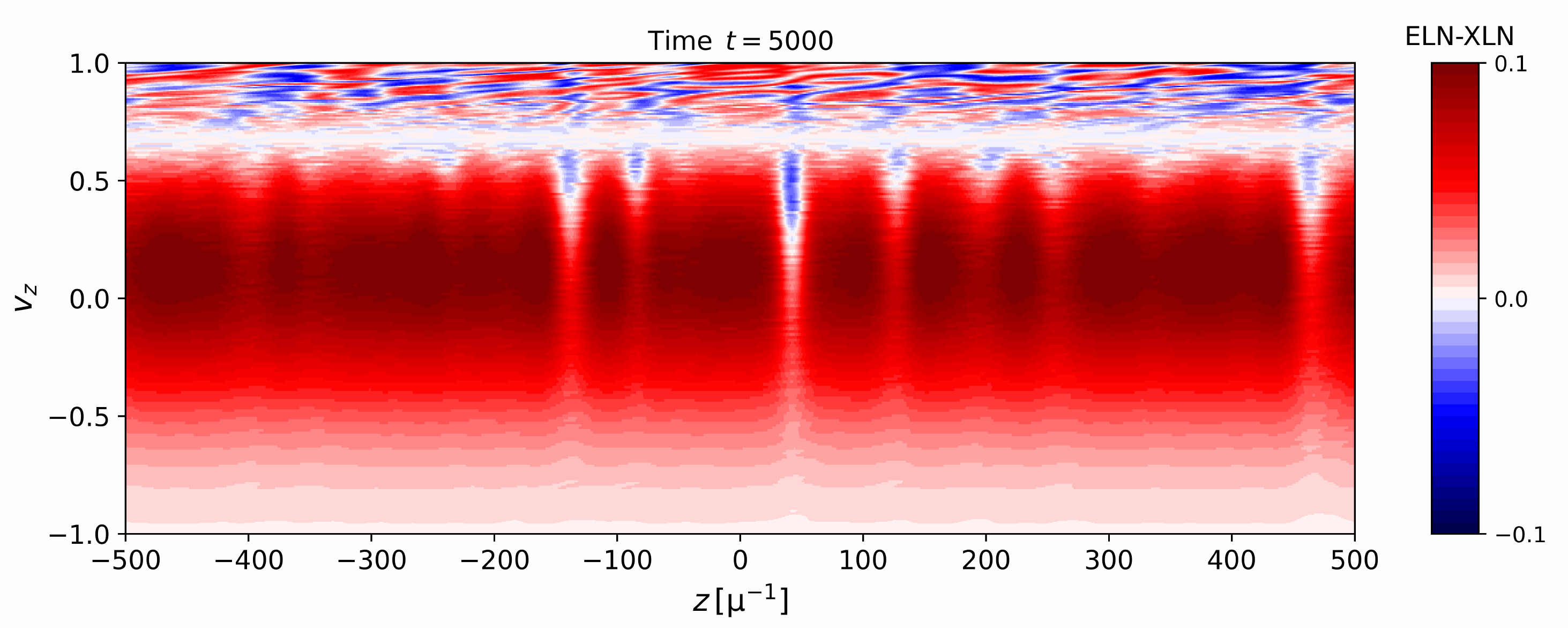}
    \includegraphics[width=1.0\linewidth]{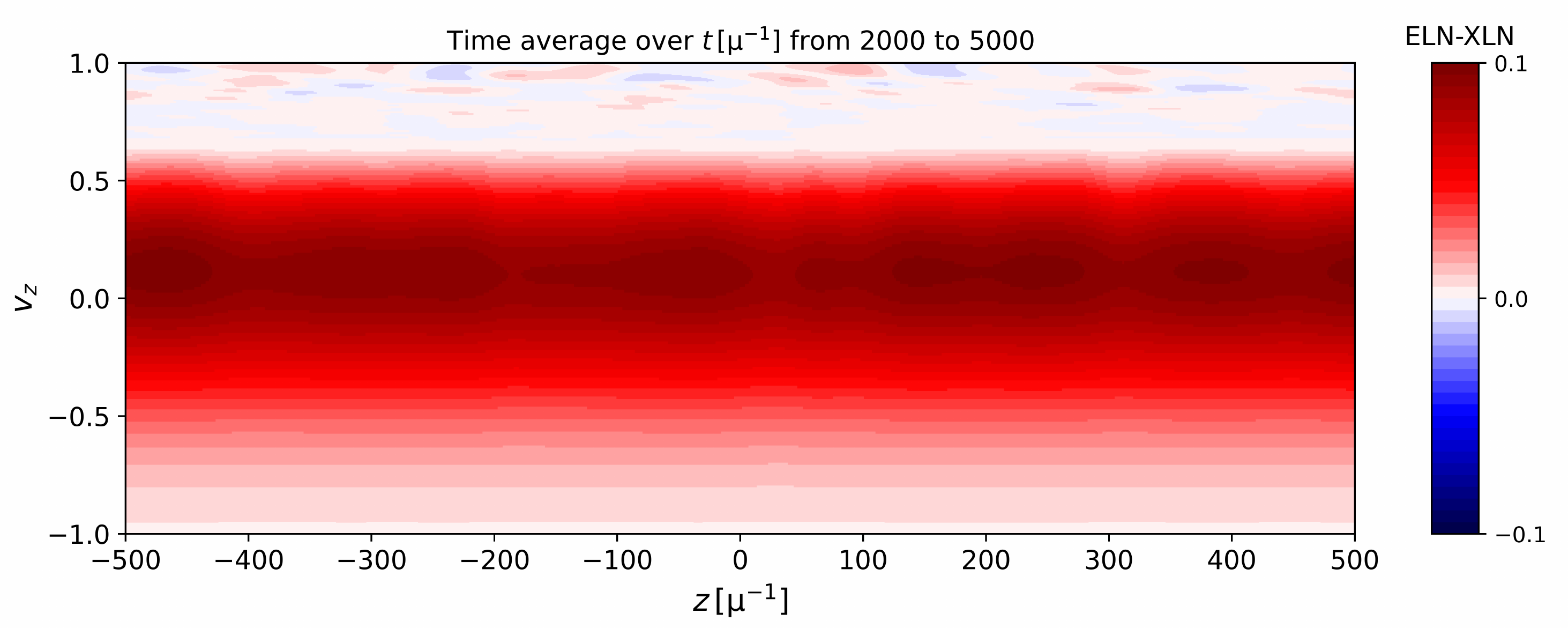}
    \caption{
    Top panel: ELN-XLN distributions as a function of $z$ and $v_z$ at the end of the simulation ($t=5000$).
    Bottom panel: same as the top panel but we take the average over time in the range of $2000 \leq t \leq 5000$.
    As shown in these panels, ELN-XLN angular crossings remain at $t=5000$, whereas the crossings in time-averaged distributions become much less prominent.
    }
    \label{fig:inhomogeneous_t_avg}
\end{figure}

It may be interesting to compare non averaged- and time-averaged angular distributions, which are displayed in Fig.\,\ref{fig:inhomogeneous_t_avg} as a function of $z$ and $v_z$.
The top panel portrays the ELN-XLN angular distributions at $t=5000$, and the bottom one shows the time-averaged distributions during the time interval of $2000 \leq t \leq 5000$.
As shown in the top panel, local ELN-XLN angular crossings remain even after the system settled into a quasi-steady state.
This is attributed to large fluctuations in space with small-scale structures, in particular, at the angular region of $v_z\gtrsim 0.7$.
On the other hand, the time-averaged ELN-XLN angular distributions smear out such local fluctuations and lead to a similar result as spatial-averaged distributions.
Indeed, the ELN-XLN becomes almost zero in $v_z\gtrsim 0.7$, illustrating the disappearance of ELN-XLN angular crossing.
This is consistent with our previous study \cite{Nagakura:2022a,Nagakura:2022b}.
Our results suggest that FFC evolves toward eliminating the angular crossings in the ELN-XLN distribution, and the system establishes a quasi-steady state with saturating flavor conversion when the crossings disappear.

\begin{figure}[t]
    \centering
    \includegraphics[width=0.9\linewidth]{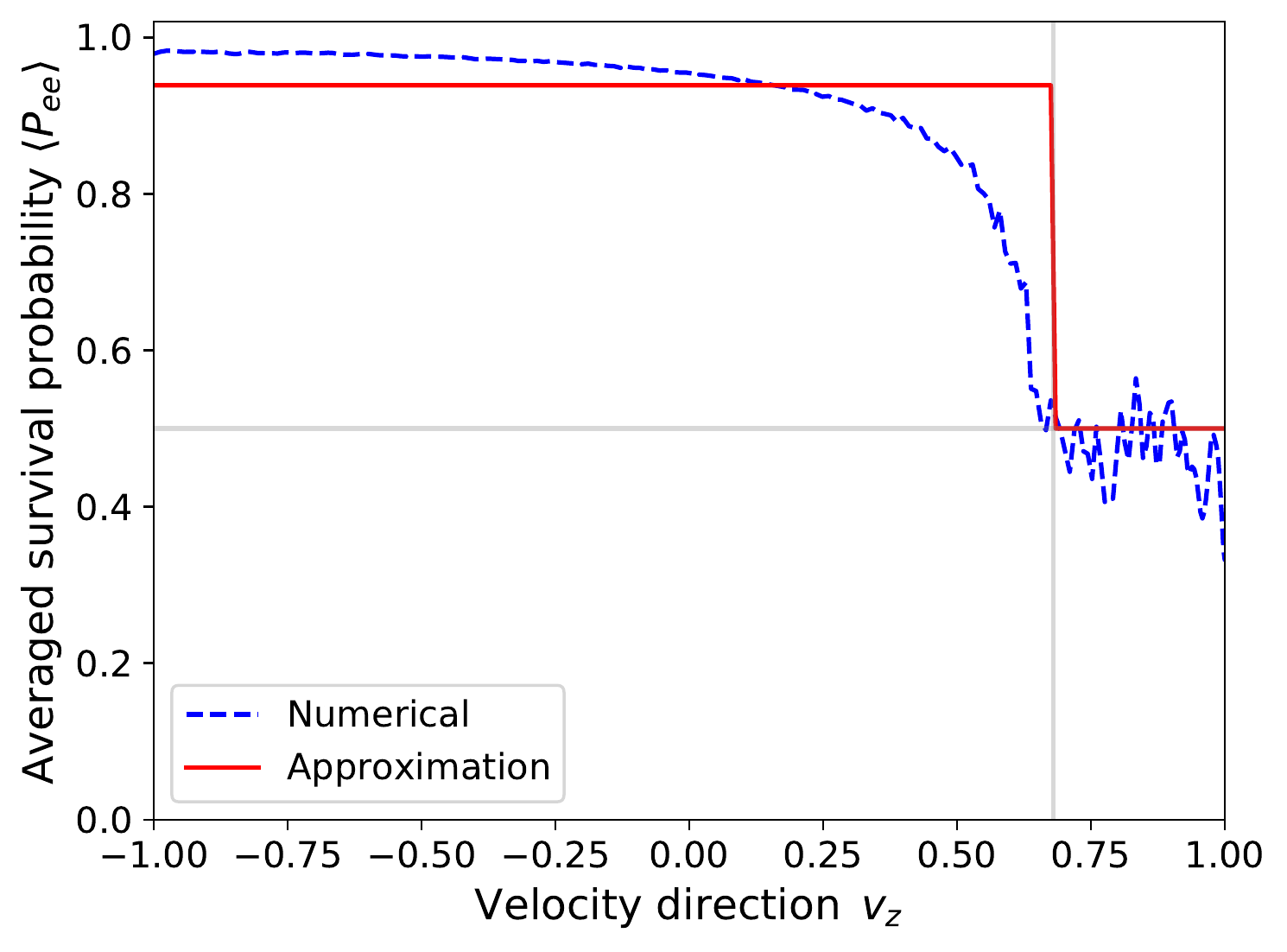}
    \includegraphics[width=0.9\linewidth]{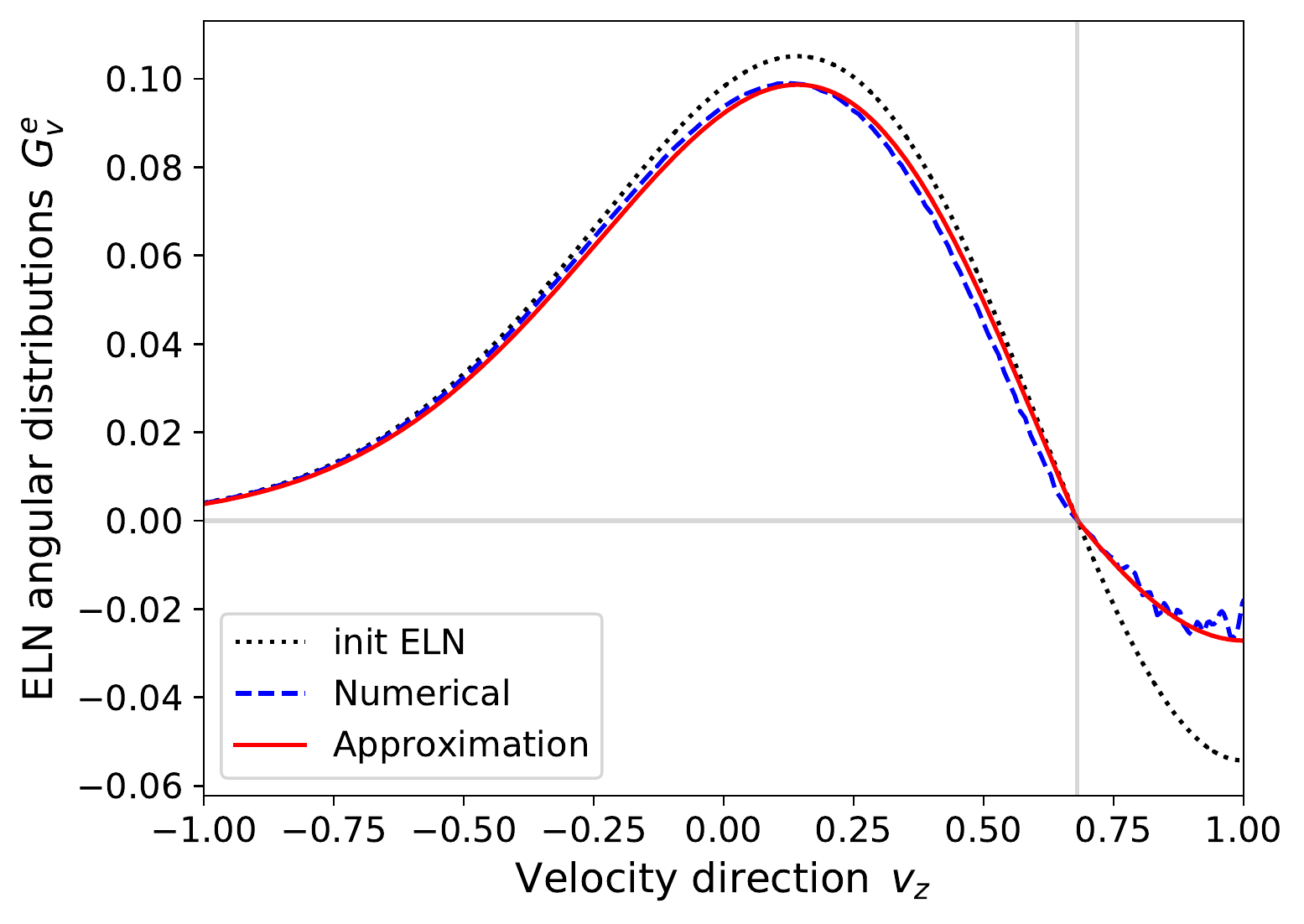}
    \caption{
    Top: Spatial-averaged survival probability at $t=5000$.
    Blue dashed line is for numerical simulation and red solid one is for our analytical scheme in Eq.\,\eqref{eq:surv_approx1}.
    Horizontal thin line corresponds to a flavor equipartition $P_{ee}=0.5$ and vertical one to the directions of an angular crossing.
    Bottom: ELN angular distributions averaged over space at $t=5000$.
    Black dotted (blue dashed) lines are initial (final) ELN angular distributions and red solid one is from our analytical scheme in Eq.\,\eqref{eq:surv_approx1}.
    }
    \label{fig:surv_approx}
\end{figure}

Finally, we apply our approximate scheme Eq.\,\eqref{eq:surv_approx1} to obtain spatial-averaged survival probability after a non-linear saturation.
Figure\,\ref{fig:surv_approx} compares the numerical results to our analytical scheme.
In the top panel, our model only broadly captures its characteristics, particularly around the crossing, because we disregard the detailed angular distributions for simplicity.
Meanwhile, we find that despite such a rough model, our approximate ELN angular distribution reproduces the numerical one well.
This is because the contribution from the compensation for the flavor equipartition is relatively small in the positive ELN parts.
This is also because the ELN is close to zero near the crossing where the deviation from numerical simulations is the greatest.
Consequently, our proposed scheme is very concise and in reasonable agreement with numerical simulations.

\subsection{Species analysis}\label{Sec.IIIC:species}
In the previous section, we have discussed the final fate of FFC in terms of ELN and XLN.
In the fast limit, the asymptotic behaviors are determined only by the ELN-XLN angular distribution.
On the other hand, the descriptions only give quantities about the difference between neutrinos and antineutrinos, not the number density or the mixing degree for each species.
In the context of CCSNe and BNSMs,
the number densities for $\nu_e$ and $\bar{\nu}_e$ are crucial to incorporate the feedback to neutrino transport, and the species-dependent neutrino angular distributions must be considered.

\begin{figure}[t]
    \centering
    \includegraphics[width=0.9\linewidth]{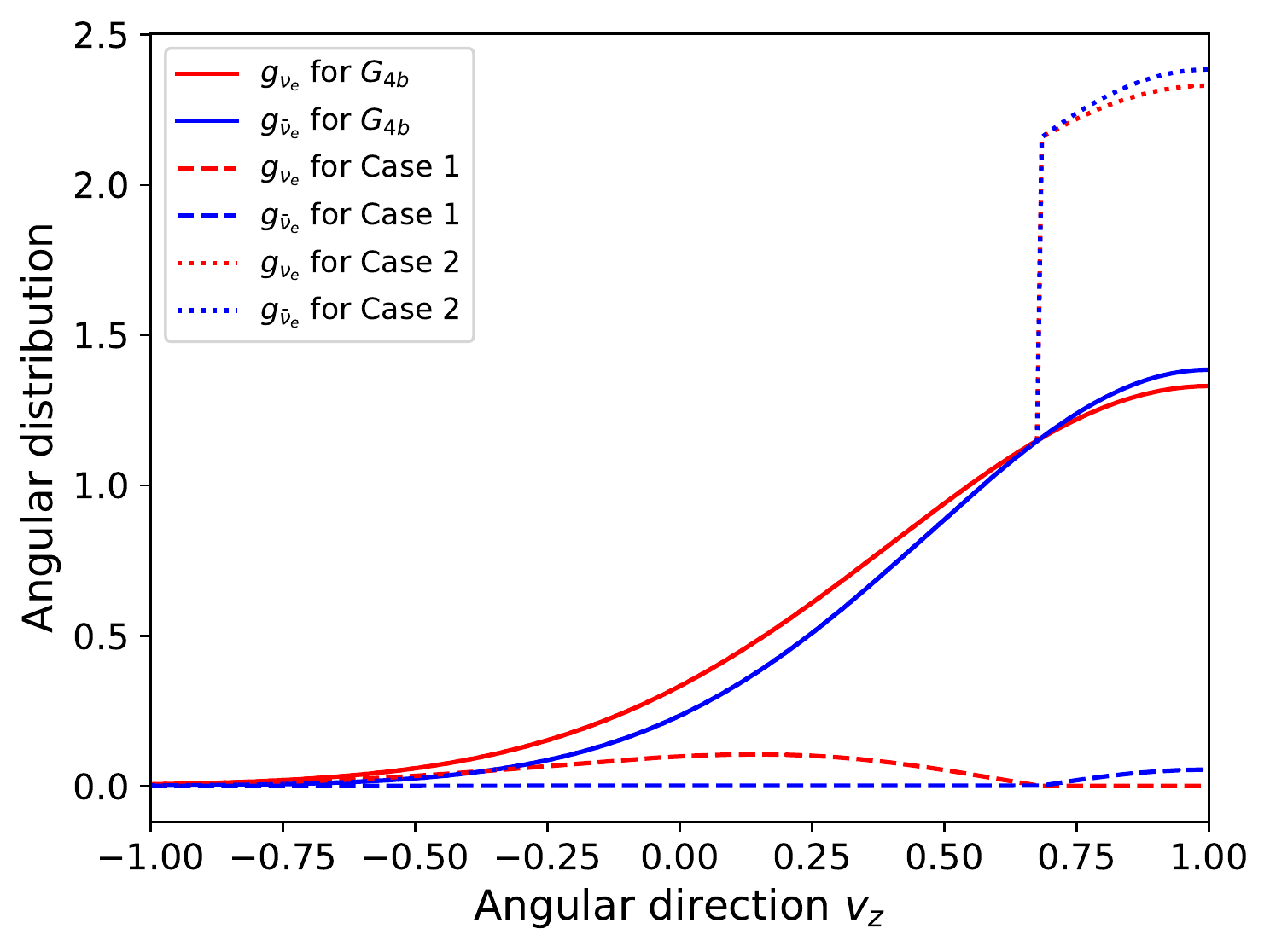}
    \includegraphics[width=0.9\linewidth]{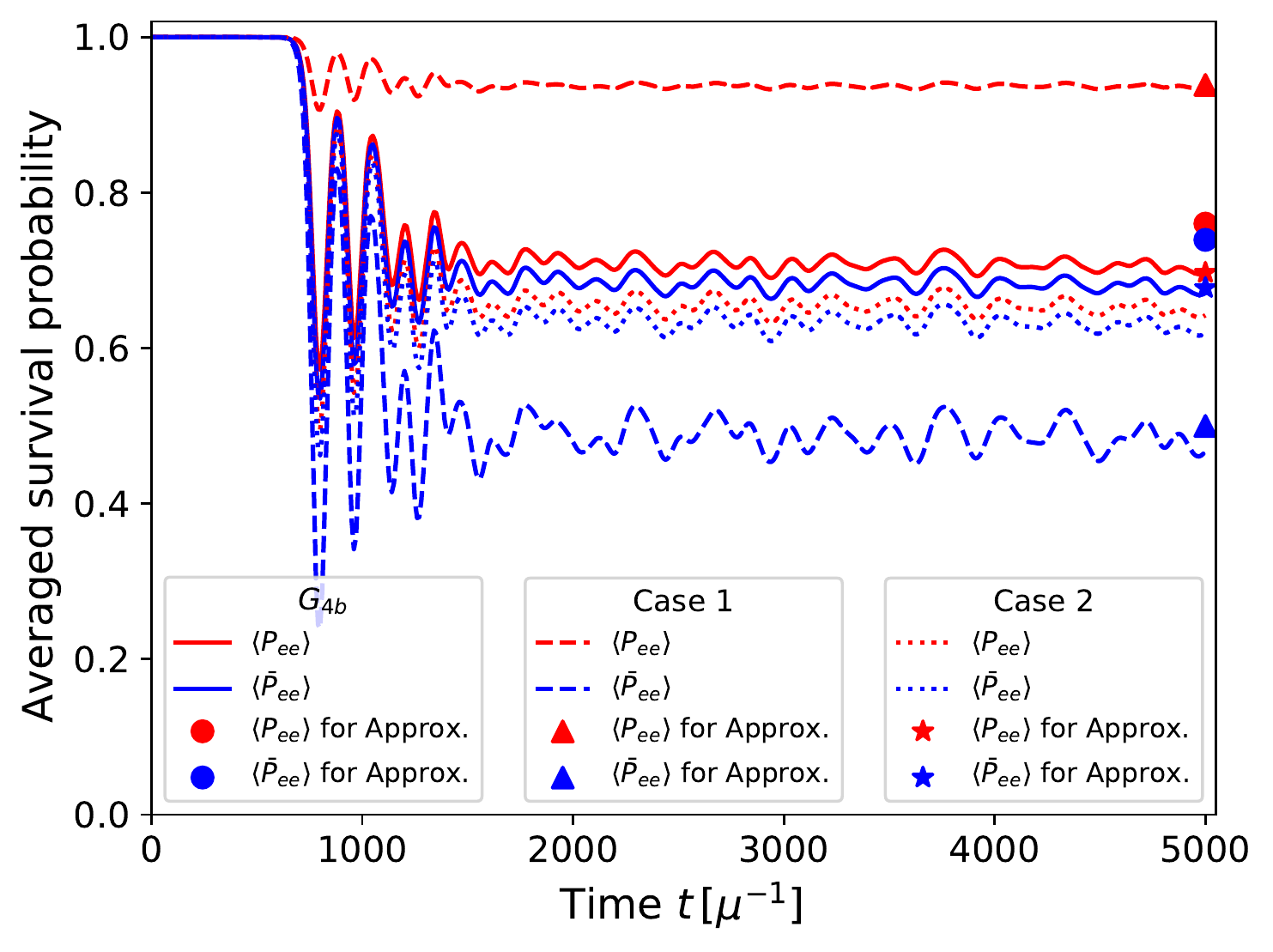}
    \caption{
    Top: angular distributions of electron (anti-)neutrinos for three cases.
    Solid red- (blue-)colored lines are angular distributions for $G_{4b}$.
    Dashed and dotted ones for Case 1 and Case 2, respectively.
    Each case generates the identical ELN angular distribution and induces the same FFC.
    Bottom: Time evolution of survival probabilities of electron (anti-)neutrino averaged over the spatial domain and the angular distributions $g_{\nu_{\alpha}}$ for three cases.
    Quite different conversion probabilities are exhibited.
    Circle, triangle, and star markers at $t=5000$ correspond to those estimated by our analytical scheme in Eq.\,\eqref{eq:surv_approx1} for each case.
    }
    \label{fig:species}
\end{figure}

Here, we discuss angular-averaged flavor conversion on each neutrino species by using the angular distributions $g_{\nu_{\alpha}}$ in Eq.\,\eqref{eq:g_nu}, not the ELN angular distributions $G_v^{e}$.
Solid lines in the bottom panel in Fig.\,\ref{fig:species} exhibit the time evolution of survival probabilities of electron (anti-)neutrinos averaged over space and neutrino angle.
This illustrates that the system reaches a non-linear saturation around $t=2000$ as seen in Fig.\,\ref{fig:inhomogeneous_P3}.
It should be mentioned that the actual steady state is never achieved, exhibiting that FFC involves stochastic fluctuations similar to turbulence.
Also, we find that the averaged survival probability for $\bar{\nu}_e$ is slightly smaller than that for $\nu_e$.
The difference hinges on the species dependence of angular distributions.
Since $\bar{\nu}_e$ exceeds $\nu_e$ in the angular regions where FFC establishes flavor equipartition, $\bar{\nu}_e$ is more converted into the heavy-leptonic flavors compared than $\nu_e$ consequently.
It means that while ELN-XLN angular distributions are sufficient to determine the dynamics of FFC, investigating the mixing degree requires species-dependent information.

Also, we plot averaged survival probabilities for our analytical model in Eq,\,\eqref{eq:surv_approx1} by two dots on $t=5000$ in the bottom panel in Fig.\,\ref{fig:species}.
Our approximate scheme exhibits relatively close but slightly larger values compared than numerical results.
The deviation comes from rough estimation near the angular crossing as shown in Fig.\,\ref{fig:surv_approx}.
For species-dependent angular distributions, the crossing is near the forward peak, so our analytical model tends to underestimate the conversion degree.
Meanwhile, in terms of the ELN, the deviations in the neutrino and antineutrino sectors cancel each other and are reduced to negligible.
This comparison lends confidence to our approximate scheme.

To delve further into the understanding of the asymptotic state of FFC, we focus on the species dependence of flavor conversion.
FFC uniquely determines the time evolution of ELN and XLN angular distributions in the fast limit.
It implies that as long as ELN-XLN angular distribution is identical, neutrinos with different species-dependent ones should experience the same flavor conversions.
Hence, we consider the extreme cases with the same ELN angular distributions as $G_{4b}$ below.

For the original ELN angular distribution in Fig.\,\ref{fig:inhomogeneous_sp_avg}, it is possible to treat the positive ELN part ($v_z\lesssim0.7$) as carried only by $\nu_e$ and the negative ELN part ($v_z\gtrsim0.7$) as carried only by $\bar{\nu}_e$; we call it Case 1.
Also, we can create another model which is bulked out only inside the angular crossing in the species-dependent angular distributions $g_{\nu}$ of $G_{4b}$; we call it Case 2.
The top panel of Fig.\,\ref{fig:species} shows the angular distributions of $\nu_e$ and $\bar{\nu}_e$ for three cases.
$G_{4b}$ (solid), Case 1 (dashed), and Case 2 (dotted) all generate identical ELN angular distributions and induce the same FFC.
On the other hand, since the fraction of number density distributed inside the angular crossing is distinct in each case, angle-averaged survival probability should exhibit different asymptotic values.
In the bottom panel of Fig.\,\ref{fig:species}, for Case 1, only $\bar{\nu}_e$ reaches the flavor equipartition within the angular crossing, and the conversion probability largely deviates that for $\nu_e$.
For Case 2, since the angular distributions are more forward-peaked, averaged survival probabilities are closer to the flavor equipartition $\langle P_{ee}\rangle=0.5$ than for $G_{4b}$.
We also plot averaged survival probabilities computed by our analytical model for both Case 1 (triangles) and Case 2 (stars) in the bottom panel.
The errors to the result of numerical simulations are within a few percent for all cases (the greatest is $\sim 9\%$ for $\langle \bar{P}_{ee}\rangle$ in $G_{4b}$).
This comparison clearly shows that our approximate scheme has the ability to capture species-dependent features.

Note that the angular distributions for Case 1 and Case 2 are not normalized in Fig.\,\ref{fig:species}.
Thereby, even though the transition probability of $\bar{\nu}_e$ for Case 1 is the largest, the converted number density of $\bar{\nu}_e$ for Case 2 is the largest.
Such behaviors imply that the depth and the width of ELN-XLN angular crossings are not directly involved in the averaged survival probability and the converted number density.
Suppose the angular direction where the flavor equipartition is achieved covers the peak of angular distributions.
In that case, electron-type neutrinos are well converted into heavy-leptonic flavors even if the angular crossing is shallow.
Generically, angular crossings generated in realistic CCSNe are relatively shallow or narrow, though FFC possibly essentially alters the neutrino number density.

\section{Comparison with previous works}
\label{Sec.IV:comparison}
It is worthwhile comparing our results to other related studies determining asymptotic states of FFC.
In this section, we describe thoroughly the similarity, difference, and novelty of our findings from these pioneering works.

In our simulations, we find that FFC in the non-linear phase undergoes a non-linear cascade into small-scale structures, similar as pointed out by Ref.\,\cite{Johns:2020a}, and then the system reaches a quasi-steady state.
Flavor equipartition is nearly achieved but it does not occur across the entire angular region.
This trend has already been reported in the previous works \cite{Bhattacharyya:2020,Bhattacharyya:2021,Wu:2021,Richers:2021,Richers:2021a,Bhattacharyya:2022}.
Parametric study in Ref.\,\cite{Wu:2021} suggested that the flavor equilibrium is a common phenomenon in FFC and neutrinos in the other side of ELN angular distribution would experience less flavor conversions constrained by the conservation laws.
The authors in Refs.\,\cite{Bhattacharyya:2020,Bhattacharyya:2021,Bhattacharyya:2022} further considered to determine such a complex asymptotic state by developing a phenomenological approach with flavor pendulum model.
They showed that the angular distributions obtained by their phenomenological models are in reasonable agreement with numerical simulations.

Although our phenomenological approach looks very similar to those proposed in the previous studies at first glance, there are some clear differences.
The most noticeable difference is that our approach does not use the flavor pendulum model.
As is well known, there is an analogy between pendulum and FFC systems (see, e.g., Ref.\,\cite{Johns:2020}), but this similarity is only guaranteed in the homogeneous FFC.
In the previous studies, they alleviate this limitation in inhomogeneous system by taking spatially-averaged neutrino distributions.
This prescription makes the inhomogeneous QKE to be similar (but not exactly the same) as the homogeneous one.

It should be stressed that the pendulum model is not always appropriate in inhomogeneous cases; more specifically, the applicability hinges on the boundary condition in space.
As can be seen from Eq.\,\eqref{eq:QKEmom}, the spatially-integrated QKE has, in general, the flux components at the boundaries, exhibiting that the obtained QKE does not become the similar form as the homogeneous one.
It is also worthy of note that the lepton number of each flavor of neutrinos can change through the neutrino flux at the boundary in general.
This exhibits another strong limitation in the pendulum model.
Since the previous studies employed a periodic boundary condition in all numerical simulations, the flux term can be dropped, and consequently their phenomenological models are in reasonably agreement with numerical simulations.
The similarity to the other previous studies employing the periodic boundary condition \cite{Wu:2021,Richers:2021,Richers:2021a,Bhattacharyya:2022} is also yielded by the conservation laws and the presence of the flux term induces a different quasi-steady state.

Our present study reveals this important limitation of pendulum model to determine asymptotic states.
The limitation indicates that we need to update the phenomenological model by different approaches for more general cases.
In this paper, we make a statement that the disappearance of ELN-XLN angular crossing corresponds to a more fundamental condition than pendulum model.
Indeed, we have witnessed in a series of our previous studies that ELN-XLN angular crossings disappear or become very shallow after the system reaches the non-linear saturation even for non-periodic boundary conditions \cite{Nagakura:2022a,Nagakura:2022b}.
One thing we need to mention here is that, although our simulations lend confidence to the statement that ELN-XLN angular crossings are fundamental quantities to characterize asymptotic states, we did not provide a rationale behind the conclusion in the previous studies.
In this paper, we address this issue through stability analysis in non-linear regime.

Stability analysis has been conducted thus far mainly to evaluate the growth rates in the linear phase.
Different from the linear phase, we leave the diagonal term $s_v$ in the governing equation for the off-diagonal term $S_v$ and try to extend the stability analysis to the non-linear regime.
On the other hand, it requires to calculate the non-linear mode couplings between $S_v$ and $s_v$ as a convolution similarly to Eq.\,\eqref{eq:QKE_fft}.
Hence, we propose a prescription that we focus on spatially- or temporally-averaged angular distributions to characterize the overall trend.
This helps us to capture the non-linear saturation in the entire system, and also guarantees the strategy that we leave aside spatial structures in our numerical simulation and focus only on the averaged domain.
It should be stressed that this stability analysis can be applied even in the case with non-periodic boundary conditions.

Aside from the discussion of ELN-XLN angular crossing, we provide another new insight regarding asymptotic states of FFC in the present study: neutrinos- and antineutrinos dependent features.
In general, we do not need to distinguish neutrinos and antineutrinos in FFC; in another words, ELN and XLN fully characterize the FFC dynamics.
For this reason, most previous studies considered the asymptotic states through ELN and XLN distributions.
Different from these studies, we pay an attention to neutrinos and antineutrinos dependent features in the asymptotic state, and some intriguing features emerge.
One of the most striking results is that shallow and narrow ELN angular crossings have the ability to mix neutrinos substantially (see Sec.\,\ref{Sec.IIIC:species} for more details).
This is attributed to the fact that ELN (or XLN) does not represent how many neutrinos and antineutrinos reside, since the number of neutrinos and antineutrinos cancel out in the computation of those for ELN and XLN.
We present in our numerical simulations that electron neutrinos with shallow ELN angular crossings can be significantly converted into heavy-leptonic flavors in angular directions where a flavor equipartition is established.
Our present study clearly exhibits that the total amount of flavor conversion depends on the angular distribution for each species, not the neutrino-flavor lepton number distribution.
It should also be emphasized that this conclusion is clearly different from other previous works, in which shallow or narrow angular crossings lead to only minimal flavor conversions \cite{Abbar:2022,Richers:2022a}.
These findings are important indications in the contexts of astrophysics, since neutrino dynamics including feedback to matter in CCSN and BNSMs are not determined by ELN (and XLN) but by species-dependent properties.
In our phenomenological model, we distinguish neutrinos and antineutrinos, which is also another noticeable difference from previous studies.

\section{Conclusions}\label{Sec.V:conclusion}
In this paper, we have presented asymptotic states of fast neutrino-flavor conversion (FFC) in local simulations with a periodic boundary condition.
Flavor instability exponentially grows by following the dispersion relation in the linear phase, and FFC occurs, gradually establishing a quasi-steady state.
We have investigated what determines the non-linear saturation in terms of the stability analysis and the conservation law for electron neutrino-lepton number (ELN) and heavy one (XLN).

Our numerical simulations show that the system reaches a quasi-steady state when the averaged ELN-XLN angular distribution has no crossings.
It is consistent with our stability analysis, suggesting that our proposed method is beneficial for analyzing the spatial- and time-averaged properties of flavor conversions.
It is worth noting that ELN-XLN angular crossings exist instantaneously, indicating that the flavor conversion would not be saturated locally.
It is also consistent with the results of our simulations, in which the exact steady state is not established, and small-scale modulations still occur within the crossing.
Our stability analysis illustrates the importance of ELN-XLN angular distributions to characterize the dynamics of fast flavor conversion even in the non-linear phase, and we conclude that the system evolves toward eliminating the ELN-XLN angular crossings (not those of ELN).

Also, we have proposed an analytic scheme determining the survival probability of each neutrino species.
Our approximate scheme is constructed to satisfy two requirements: the disappearance of ELN-XLN angular crossings (a flavor equipartition) and the conservation laws for ELN (XLN).
We demonstrated that the overall trend of FFCs in asymptotic states can be qualitatively determined by the two conditions, and we provided the numerical recipe as an approximate scheme (see Eqs.\,\eqref{eq:surv_approx1} and \eqref{eq:surv_approx2}).

We have exhibited the averaged survival probabilities for angular distributions different in electron-type neutrinos but with identical ELN ones.
ELN-XLN angular distributions characterize the dynamics of FFC and uniquely determine the quasi-steady state.
On the other hand, the conversion degree for each species is mainly determined by how many neutrinos are distributed in the directions where a flavor conversion, particularly a flavor equipartition, is induced.
Our results emphasize the importance of species-dependent information for the total amount of flavor conversion.
Even if the angular crossings are shallow or narrow, they can significantly impact on the flavor contents through the flavor equipartition in the angular directions where neutrinos are more abundant.

Although our presented descriptions for the non-linear phase of FFC are very clear and it will be useful to develop the sub-grid model of FFC, there remain some crucial issues.
As discussed in the present study, the boundary condition is a key ingredient to characterize the asymptotic state, but it is a non-trivial issue which boundary condition (periodic or Dirichlet) is appropriate.
This would hinge on both global and local properties of neutrino radiation fields and needs to be investigated in detail.
It is interesting to note that the quasi-steady state with a periodic boundary condition temporarily appeared in the early phase of FFC simulations in our previous study \cite{Nagakura:2022b}, in which we employed the Dirichlet boundary condition.
We witnessed that such a quasi-steady state is sustained up to the time when the neutrinos emitted from the inner boundary reach there, but it transits to another asymptotic state at the end of the simulation.
This suggests that we need to consider the asymptotic state of FFC with arbitrary boundary conditions.
We leave its detailed study to future work.
Another crucial issue would be the impact of collisions;
indeed, FFC could be qualitatively different due to the multi-energy effects of neutrino-matter interactions \cite{Kato:2022}.
The asymptotic state of FFC with collisions would be considered by correcting Eq.\,\eqref{eq:QKEmom} (adding the source term and restoring the energy-dependence in QKE); the detailed study is also currently underway. 
As such, there are many works to be needed, but the study of the asymptotic state, as done in the present study, will shed light on key ingredients characterizing flavor conversion even in such complex systems.

\begin{acknowledgments}
We are grateful to Shoichi Yamada, Chinami Kato, and Tomoya Takiwaki for useful comments and discussions.
MZ is supported by the Japan Society for Promotion of Science (JSPS) Grant-in-Aid for JSPS Fellows (Grants No. 22J00440) from the Ministry of Education, Culture, Sports, Science, and Technology (MEXT) in Japan.
The numerical computations were carried out on Cray XC50 at the Center for Computational Astrophysics, National Astronomical Observatory of Japan.
\end{acknowledgments}


\bibliographystyle{apsrev4-1}
\bibliography{Nonlinear_stability}

\end{document}